\documentclass[epj]{svjour}
\usepackage{graphics}
\usepackage{color}
\begin{document}
\title{A note on the electrochemical nature of the thermoelectric power}
\author{Y. Apertet\inst{1}\thanks{\emph{Email address:} yann.apertet@gmail.com} \and H. Ouerdane\inst{2} \inst{3} \and C. Goupil\inst{4} \and Ph. Lecoeur\inst{5}
}%
%
\offprints{Y. Apertet}          
\institute{Lyc\'ee Jacques Pr\'evert, F-27500 Pont-Audemer, France \and Russian Quantum Center, 100 Novaya Street, Skolkovo, Moscow region 143025, Russian Federation \and UFR Langues Vivantes Etrang\`eres, Universit\'e de Caen Normandie, Esplanade de la Paix 14032 Caen, France \and Laboratoire Interdisciplinaire des Energies de Demain (LIED), UMR 8236 Universit\'e Paris Diderot, CNRS, 5 Rue Thomas Mann, 75013 Paris, France \and Institut d'Electronique Fondamentale, Universit\'e Paris-Sud, CNRS, UMR 8622, F-91405 Orsay, France}
\date{Received: date / Revised version: date}
%
\abstract{
While thermoelectric transport theory is well established and widely applied, it is not always clear in the literature whether the Seebeck coefficient, which is a measure of the strength of the mutual interaction between electric charge transport and heat transport, is to be related to the gradient of the system's chemical potential or to the gradient of its electrochemical potential. The present article aims to clarify the thermodynamic definition of the thermoelectric coupling. First, we recall how the Seebeck coefficient is experimentally determined. We then turn to the analysis of the relationship between the thermoelectric power and the relevant potentials in the thermoelectric system: As the definitions of the chemical and electrochemical potentials are clarified, we show that, with a proper consideration of each potential, one may derive the Seebeck coefficient of a non-degenerate semiconductor without the need to introduce a contact potential as seen sometimes in the literature. Furthermore, we demonstrate that the phenomenological expression of the electrical current resulting from thermoelectric effects may be directly obtained from the drift-diffusion equation.
} 
\maketitle
\section{Introduction}
Thermoelectricity is a mature yet still very active area of research covering various fields of physics, physical chemistry, and engineering. The large interest in thermoelectric systems is mostly due to the promising applications in the field of electrical power production from waste heat as thermoelectric devices may be designed for specific purposes involving powers over a range spanning ten orders of magnitude: typically from microwatts to several kilowatts. Further, thermoelectricity also provides model systems that are extremely useful in the development of theories in irreversible thermodynamics \cite{deGroot,Apertet2012}.

The discovery of the thermoelectric effect is usually attributed to Seebeck. In 1821, he published the results and analysis of his experiments aiming at establishing a magnetic polarization in a metallic circuit simply by perturbing the thermal equilibrium across this latter \cite{Seebeck1821}. More precisely, Seebeck described the appearance of a magnetic field within a closed electrical circuit made of two dissimilar materials as the junctions between these materials were maintained at different temperatures. While Seebeck interpreted the observed phenomenon as a thermomagnetic effect, Oersted soon reexamined Seebeck's work and showed that in this case the magnetic field was an indirect effect as it originated in the presence of an electromotive force induced by the temperature difference \cite{Oersted1823}. The proportionality coefficient between this electromotive force and the temperature difference across the system is the thermoelectric power, which has also been coined as ``Seebeck coefficient''.

The definition of the thermoelectric coupling has later been extended from that derived from the first experiments to both thermodynamic \cite{Callen1948} and microscopic \cite{Herring1954,Price1956,Heikes1961,Wood1988} properties of materials. However, as of yet, there still is no clear consensus on its relationship with the various thermodynamic potentials and their variations (see, e.g., refs. \cite{Kittel,Mahan2000,Cai2006,Peterson2010,Varlamov2013,Ouerdane2015,Behnia2015}). Indeed as the terminology and conventions may vary from a discipline to another, say, e.g., solid-state physics and electrochemistry, it is not always straightforward to establish a clear distinction or relevant associations between Fermi energy at zero or finite temperature, electrochemical potential, voltage, Fermi level relative either to the conduction band minimum or to the vacuum, and chemical potential.

In this article, we discuss the definition of the Seebeck coefficient focusing particularly on the distinction between chemical and electrochemical potentials. First, in sect.~\ref{exp}, we address the experimental determination of the Seebeck coefficient in order to identify the quantities of interest. Next, the purpose of sect.~\ref{link} is to demonstrate that a clear physical picture of thermoelectric phenomena at the microscopic scale may be obtained on the condition that the potentials are carefully introduced. For this purpose, we review the standard definitions given in the literature to remove any confusion between the chemical and electrochemical potentials before we present and discuss our derivation of the Seebeck coefficient for a non-degenerate semiconductor. 

\section{\label{exp} Experimental determination of the thermoelectric power}
\begin{figure}
	\centering
	\resizebox{0.5\textwidth}{!}{
		\includegraphics{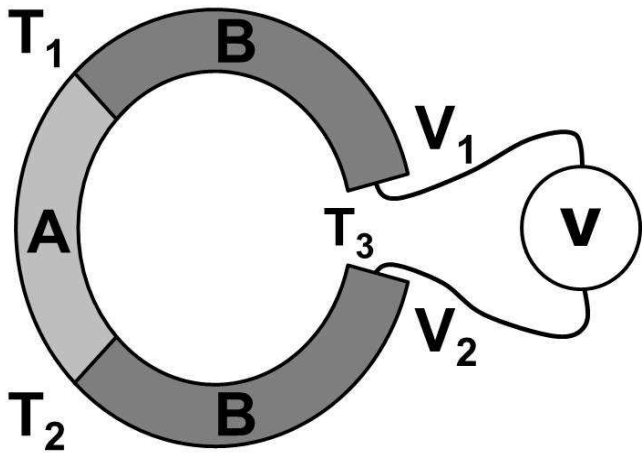}}
	\caption{Determination of the Seebeck coefficient for a circuit composed of two dissimilar materials.}
	\label{fig:figure1}
\end{figure}

The determination of the Seebeck coefficient traditionally involves components made of dissimilar materials, which we label A and B respectively. The two materials are combined to obtain two junctions as depicted in fig. \ref{fig:figure1}. These junctions are then brought to different temperatures $T_1$ and $T_2$. An isothermal voltage measurement at a temperature $T_3$, is performed between the free ends of the component B. The voltage thus measured is $V_2 - V_1$ (this notation allows to clearly define a direction for the voltage) and the Seebeck coefficient $\alpha_{\rm AB}$ associated with the global system, i.e. the couple AB, is defined as the proportionality coefficient between the resulting voltage and the applied temperature difference:
\begin{equation}
\alpha_{\rm AB} = \frac{V_2 - V_1}{T_2 - T_1}.
\end{equation}

\noindent The coefficient $\alpha_{\rm AB}$, obtained for the whole circuit, is related to the Seebeck coefficient of each material through \cite{Callen1985}:
\begin{equation}
\alpha_{\rm AB} = \alpha_{\rm B} -\alpha_{\rm A},
\end{equation}
\noindent where $\alpha_{\rm A}$ and $\alpha_{\rm B}$ are the Seebeck coefficients of the materials A and B respectively.

From an experimental viewpoint, the presence of the material B ($\neq$ A) is mandatory as it is associated with the probe's wires (see, e.g., ref. \cite{Zhou2005}). However, if its Seebeck coefficient $\alpha_{\rm B}$ is sufficiently small to be neglected, the measurement may be used to determine directly the Seebeck coefficient of material A. In this case, one gets:
\begin{equation} \label{def1}
\alpha_{\rm A} = - \frac{V_2 - V_1}{T_2 - T_1}.
\end{equation}
\noindent Note the presence of a minus sign in the expression above: It is often overlooked in the literature but, fortunately, that omission is most of the time compensated by the absence of a clear sign convention for the measured voltage.

Let us now turn to the analysis of the measured quantities. While the temperature is not subject to questioning, the voltage obtained from a voltmeter must be defined unambiguously. Indeed, it appears that its connection to the microscopic and thermodynamic properties of materials has remained unclear for quite some time, leading Riess to publish in 1997, hence fairly recently, an article untitled ``\emph{What does a voltmeter measure ?}'' \cite{Riess1997}. In that paper, Riess demonstrated that the voltage measured by a voltmeter between two points in a circuit is the difference of electrochemical potentials $\widetilde{\mu}$ at the two considered points divided by the elementary electric charge $e$, but \emph{not} the difference between the electrostatic potentials $\varphi$ alone. The potential $V$ might thus be defined as $V = - \widetilde{\mu}/e$. This result is recovered when one measures the voltage at the ends of a pn junction at equilibrium: While there is a built-in electric field associated with the depletion layer, the measured voltage remains zero.  The Seebeck coefficient thus appears as a link between the applied temperature difference and the resulting \emph{difference of electrochemical potential} between the two junctions. 

The simple technique presented here is not the only one used to determine the thermoelectric power of a given material. Indeed, since the measurement always involves a couple of materials, the absolute Seebeck coefficient of the second material has to be known accurately. To obtain this value, it is possible to use low temperature measurement to reach superconducting state where $\alpha = 0$ and then derive higher temperatures values using the Thomson coefficient that can be measured for a single material. For a detailed presentation of the Seebeck coefficient metrology, the reader may refer to the instructive review by Martin \emph{et al}. \cite{Martin2010}.

\section{\label{link}Relationship between the thermoelectric power and the electrochemical potential}
In order to better understand the influence of each potential, we identify the respective effects of temperature bias, concentration difference, and electric charge, and we discuss the relationship between chemical potential, electrochemical potential and the band diagram of materials. We then derive the Seebeck coefficient in the simple case of a non-degenerate semiconductor to illustrate the contribution of each potential.

\subsection{Definition of the thermopower}
The Seebeck coefficient may be obtained from a microscopic analysis of the considered materials, with the \emph{local} version of Eq.~(\ref{def1}), in open-circuit condition, i.e., with a vanishing electrical current: 

\begin{equation}\label{alpha3D}
\alpha = \frac{\nabla \widetilde{\mu}}{e \nabla T},
\end{equation}

\noindent where $\widetilde{\mu}$ and $T$ are respectively the local electrochemical potential and temperature, defined at each point of the system. The notation $\nabla$ is associated with the gradient of each quantity. In the following, for the sake of simplicity, we consider a unidimensional system so that the spatial gradient reduces to its $x$-component: $\nabla_x$.

\subsection{Distinction between the potentials}

Consider a semiconductor sample at thermal equilibrium and characterized by a spatially inhomogeneous doping. As the carrier concentration is nonuniform, a particle current takes place from the region of higher concentration to that of lower concentration: This is the diffusion process associated with the variation of the carriers' chemical potential across the system. This type of electrical current is referred to as the diffusion current. The inhomogeneous electron population in the system thus generates an electric potential difference and hence a built-in electric field which influences the electrons' motion in such a fashion that it tends to curb the diffusion current. The electron motion driven by the built-in electric field is the drift current, which, at thermal equilibrium, exactly cancels the diffusion current, in accordance with the principle of Le Chatelier and Braun. In this case, the measured voltage across the system always remains zero and there is no net electrical current even if the system is short-circuited: The electric field associated with the electrical potential variation is obviously not an electromotive field. However, if the electrons are placed in a non-equilibrium situation caused by a thermal bias applied across the system, a non vanishing electric current may be obtained when the circuit is closed. This current obviously stems from the uncompensated contributions of both the diffusion and drift of charge carriers, and it is traditionally related to the gradient of the temperature and to the gradient of the electrochemical potential.

The electrochemical potential $\widetilde{\mu}$ of a population of electrically charged particles is the sum of a chemical contribution $\mu$, the chemical potential, and of an electrical contribution $\mu_{\rm e}$ \cite{Callen1985}:

\begin{equation}\label{defCallen}
\widetilde{\mu} = \mu + \mu_{\rm e}.
\end{equation}

\noindent Note that the quantities we just referred to as potentials are actually energies. The electrical contribution $\mu_{\rm e}$ may be expressed as a function of the electrostatic potential $\varphi$ (a genuine potential contrary to $\widetilde{\mu}$ and $\mu$) so that the electrochemical potential reads:

\begin{equation} \label{bien}
\widetilde{\mu} = \mu + q\varphi,
\end{equation}

\noindent where $q$ is the electrical charge of the considered particle. When used in solid state physics, these quantities have to be related to an energy band diagram. This correspondence may be found for example in the book of Heikes and Ure \cite{Heikes1961}: Considering the example of an n-doped semiconductor, the electrochemical potential $\widetilde{\mu}$ corresponds to the Fermi level, the electrostatic energy $-e\varphi$ corresponds to the energy level of the bottom of the conduction band while the chemical potential $\mu$ corresponds to the difference between these two quantities and is often called Fermi energy
\begin{figure}
	\centering
	\resizebox{0.5\textwidth}{!}{
		\includegraphics{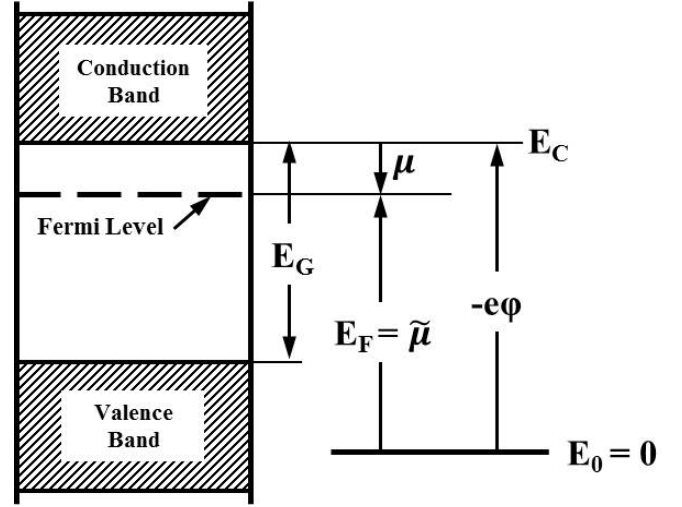}}
	\caption{Energy levels in an n-type semiconductor highlighting the notations used in this article (adapted from ref. \cite{Heikes1961}). The energy $E_{\rm G}$ refers to the bandgap energy.}
	\label{fig:figure2}
\end{figure}
These notations are summarized on fig.~(\ref{fig:figure2}). The difference between Fermi level ($\widetilde{\mu}$) and Fermi energy ($\mu$) was already highlighted by Wood \cite{Wood1988}: ``\emph{The difference between the Fermi energy and the Fermi level should be noted. The Fermi energy is generally measured from the adjacent conducting band edge (valence or conduction band for holes or electrons, respectively), i.e. a reference level which may vary in energy, whereas the Fermi level is measured from some arbitrary fixed energy level}''. This last remark stresses the importance of the choice of an energy reference, which is a key parameter: To express energies in a semiconductor, the bottom of the conduction band is often used as the reference \cite{Kittel}; however, for studies of non-equilibrium phenomena such as thermoelectricity, it is mandatory to define an arbitrary fixed energy reference independent of the position within the material since both $\widetilde{\mu}$ and $\mu$ may vary along the system. It seems the only way to correctly describe the relative displacement of these energies. Note that the vacuum level infinitely far from the system, $E_\infty$, might be a good and meaningful energy reference. 

\begin{figure}
	\centering
	\resizebox{0.5\textwidth}{!}{
		\includegraphics{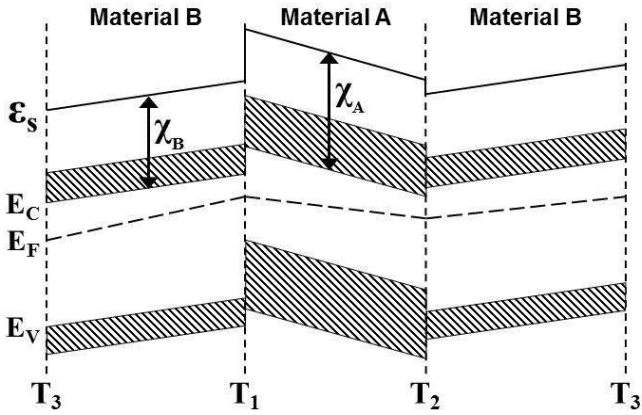}}
	\caption{Schematic illustration (adapted from Ref.~\cite{Chambers1977}) of the variations of the bottom of the conduction band, $E_C$, the top of the valence band, $E_V$, the Fermi level, $E_F$, and the vacuum level just outside the material, $\epsilon_S$, all along the circuit depicted in Fig.~\ref{fig:figure1}. The slopes of the lines have been greatly exaggerated for clarity, and band bending at the interfaces has been neglected.}
	\label{fig:figure3}
\end{figure}

Figure~\ref{fig:figure3} illustrates the variations of the different energies around the circuit depicted in Fig.~\ref{fig:figure1} in the case of semiconductor materials. It highlights the difference between the slope of the bottom of the conduction band and the slope of the Fermi level: The variation of the chemical potential thus differs from the variation of the electrochemical potential. Distinguishing these two energies is, therefore, crucial to properly evaluate the Seebeck coefficient. Figure~\ref{fig:figure3} also displays the vacuum level $\epsilon_S$ just outside the material (different from $E_\infty$). This vacuum level is related to the bottom of the conduction band through the affinity $\chi$ of the material. The discontinuities in $\epsilon_S$ at the interfaces might be seen as contact potentials. On the contrary, the Fermi level $E_F$ is continuous along the system, even at the interfaces. Its variation however undergoes a sudden change at the interface, reflecting both changes in temperature gradient (assumed constant in a given material) and in Seebeck coefficient from a material to an other. The thermopower is indeed associated with bulk material but not to interfaces. A similar figure for a system made of metals can be found in Ref.~\cite{Chambers1977}.

\subsection{From potentials to thermoelectric power: the illustrative case of a non-degenerate semiconductor}

We emphasise the importance of the distinction between $\widetilde{\mu}$ and $\mu$ on the derivation of the thermoelectric power using the example of a non-degenerate semiconductor doped with electrons. In this case, the expression of the carrier concentration $n$ is rather simple:

\begin{equation}\label{electrondensity}
n(T) = N \exp \left(\frac{\widetilde{\mu} - E_{\rm C}}{k_{\rm B} T}\right) = N \exp \left(\frac{\mu}{k_{\rm B} T}\right), 
\end{equation}

\noindent with 

\begin{equation}
N = 2\left(\frac{2\pi m_{\rm eff} k_{\rm B} T}{h^2}\right)^{3/2},
\end{equation}

\noindent and where $E_{\rm C}$ is the energy level of the bottom of the conduction band, $m_{\rm eff}$ is the electron effective mass, $k_{\rm B}$ is the Boltzmann constant and $h$ is the Planck constant. The Seebeck coefficient is associated with non-equilibrium phenomena, and, as such, it is tightly linked to transport properties of electrons inside the material. To take account of these properties, we build on the drift-diffusion equation used to obtain the net electrical current density $J_x$: 

\begin{equation}\label{eq:derivdiff}
J_x = e n M_n \mathcal{E}_x + e D_n \nabla_x n,
\end{equation}

\noindent where $M_n$ and $D_n$ are the electron mobility and diffusivity, and where the electric field $\mathcal{E}_x$ is related to the energy level $E_{\rm C}$ through: 

\begin{equation}
\mathcal{E}_x = - \frac{\nabla_x E_{\rm C}}{q} = \frac{\nabla_x E_{\rm C}}{e}.
\end{equation}

\noindent At first, we assume a situation where the electron diffusivity $D_n$ does not depend on the other parameters, including the position. The variation of $D_n$ will be discussed further below. 

The Seebeck coefficient is obtained setting $J_x = 0$. However this current density should be related first to $\nabla_x T$ and $\nabla_x \widetilde{\mu}$ rather than to $\mathcal{E}_x$ and $\nabla_x n$. To do so, we evaluate the gradient of the electron density given by eq. (\ref{electrondensity}) considering that $E_{\rm C}$, $\widetilde{\mu}$ and $T$ may vary along the material. This approach is seldom found in the literature as one often sets $E_{\rm C} = 0$, thus considering the bottom of the conduction as the reference everywhere in the nonequilibrium system. As already stressed, this viewpoint is misleading for thermoelectric phenomena. From eq.~(\ref{electrondensity}), the gradient of electron density reads: 
  
\begin{equation}\label{nablan}
\nabla_x n = \frac{3}{2} \frac{n \nabla_x T}{T} + \frac{n}{k_{\rm B} T^2} \left[T \left(\nabla_x \widetilde{\mu} - e\mathcal{E}_x\right) - \mu \nabla_x T \right]. 
\end{equation}

\noindent We then use this equality along with Einstein's relation between the electron mobility $M_n$ to the electron diffusivity $D_n$: 

\begin{equation}
\frac{M_n}{D_n} = \frac{e}{k_{\rm B} T} 
\end{equation}

\noindent to modify eq. (\ref{eq:derivdiff}) as follows:

\begin{equation}\label{eq:derivdiffTE}
J_x = e n M_n \left(\frac{\nabla_x \widetilde{\mu}}{e}  + \frac{k_{\rm B}}{e} \left[\frac{3}{2} - \frac{\mu}{k_{\rm B} T}\right] \nabla_x T \right).
\end{equation}

\noindent Now, setting $J_x = 0$ and using the definition given in eq.~(\ref{alpha3D}), we find: 

\begin{equation}\label{alphaSC}
\alpha = - \frac{k_{\rm B}}{e} \left[\frac{3}{2} - \frac{\mu}{k_{\rm B} T}\right],
\end{equation} 

\noindent with a \emph{constant} electron diffusivity, which is the expected expression for a non-degenerate semiconductor. Further, this result may also be interpreted by looking at the net thermal energy transported by each carrier transported inside the material, i.e., $q\Pi$, where $\Pi$ is the Peltier coefficient \cite{Callen1985}. For electrons, this energy is
\begin{equation}\label{totalheat}
- e \Pi = \frac{3}{2}k_B T - \mu,
\end{equation}
\noindent since it corresponds to the energy above the Fermi level $\widetilde{\mu}$ and hence to the sum of the average thermal energy for free electrons and of the energy between the Fermi level and the bottom of the conduction band, i.e., $-\mu$. We thus recover the second Kelvin relation relating the Seebeck and Peltier coefficients: $\Pi = \alpha T$.

\subsection{Taking into account diffusivity variation}

If we relax the assumption of constant diffusivity $D_n$, this latter becomes a function of the spatial coordinate $x$ and we end up with the so-called Stratton equation \cite{Lundstrom2009}: 

\begin{eqnarray}\label{eq:Stratton}
J_x &=& e n M_n \mathcal{E}_x + e \nabla_x \left(D_n n\right)\nonumber \\
&=& e n M_n \mathcal{E}_x + e D_n \nabla_x \left(n\right) + e n \nabla_x \left(D_n\right),
\end{eqnarray} 

\noindent It corresponds to a more general form of the drift-diffusion equation, which contains a third contribution to the carrier motion, directly linked to the gradient of diffusivity along the system. To evaluate its effect on the thermoelectric power, we may reexpress it as a function of the temperature gradient using the relation between the diffusivity $D_n$ and the relaxation time of the carriers $\tau$. Since $M_n = e\tau / m_{\rm eff}$, the Einstein relation reads: 

\begin{equation}\label{Dn}
D_n = k_{\rm B} T \frac{\tau}{m_{\rm eff}}.
\end{equation}

\noindent To keep the calculations on an analytical level, we assume that we deal with low-energy conduction electrons, and we express the relaxation time using a power law of the form: $\tau \propto (E-E_{\rm C})^s$, where $E$ is the total energy of the carrier and $s$ is a characteristic exponent depending on the scattering mechanisms \cite{Lundstrom2009}. Some typical values for this exponent are given in Table~\ref{table1}.
\begin{table}
	\centering
	\begin{tabular}{|l|c|}
  \hline
  Scattering mechanism & Exponent \emph{s}\\
  \hline
  Acoustic phonon & -1/2 \\
  Ionized impurity (strongly screened) & -1/2 \\
  Neutral impurity & 0 \\
  Piezoelectric & +1/2 \\
  Ionized impurity (weakly screened) & +3/2 \\
  \hline
\end{tabular}
	\caption{Values of the exponent \emph{s} for different scattering mechanisms (adapted from Ref. \cite{Lundstrom2009}).}
	\label{table1}
\end{table}
Note that the energy $E-E_{\rm C}$ corresponds to the thermal energy of the carriers in the conduction band and may thus be approximated by its average value, i.e., $3/2k_{\rm B} T$. Replacing $\tau$ in eq. (\ref{Dn}), we obtain: 

\begin{equation}\label{nablaDn}
\nabla_x \left(D_n\right) =  \frac{D_n}{T} (1+s) \nabla_x \left(T\right).
\end{equation} 

\noindent Finally, inserting eq. (\ref{nablaDn}) and eq. (\ref{nablan}) in eq. (\ref{eq:Stratton}) yields: 

\begin{equation}\label{eq:derivdiffTE2}
J_x = e n M_n \left(\frac{\nabla_x \widetilde{\mu}}{e}  + \frac{k_{\rm B}}{e} \left[\frac{5}{2} + s - \frac{\mu}{k_{\rm B} T}\right] \nabla_x T \right),
\end{equation}

\noindent and consequently:

\begin{equation}\label{alphaSC2}
\alpha =  - \frac{k_{\rm B}}{e} \left[\frac{5}{2} + s - \frac{\mu}{k_{\rm B} T}\right].
\end{equation}

\noindent The contribution of the diffusivity gradient to the thermoelectric power is $-(1+s) k_{\rm B}/e$ and hence depends only on the scattering parameter $s$. This term has also been recovered by Cai and Mahan \cite{Cai2006} using a Boltzmann equation. Note that this term was also introduced by Ioffe \cite{Ioffe1960} with the notation $\alpha_D$. However, Ioffe used a different power law: He assumed that the carrier's mean free path $\overline{l}$ is proportional to $(E-E_{\rm C})^r$. He consequently found that $\alpha_D = -(1/2+r) k_{\rm B}/e$. This discrepancy is quite easy to understand since $\tau$ is proportional to $\overline{l} /\sqrt{E-E_{\rm C}}$.

\section{Discussion}

\subsection{An unusual derivation}
While the result given in Eq.~(\ref{alphaSC}) is well-known, its derivation presented here is quite original. Indeed, it was directly obtained from the drift-diffusion equation. Thus, the phenomenological equation associated with thermoelectric transport: 

\begin{equation}\label{eq:TEmicro}
J_x = \sigma \frac{\nabla_x \widetilde{\mu}}{e} - \sigma \alpha \nabla_x T,
\end{equation}

\noindent where $\sigma = e n M_n$ is the electrical conductivity, is identical to eq. (\ref{eq:derivdiffTE2}) (or to eq. (\ref{eq:derivdiffTE}) depending on the hypothesis made). This latter appears as a modified form of the drift-diffusion equation, which accounts for the couple of variables [$\widetilde{\mu}$, T], or more precisely their gradients, rather than the traditional couple [n, $\varphi$]. This modification puts forth the fact that the first term of the right hand side of eq. (\ref{eq:TEmicro}) does not correspond any longer to the genuine local form of Ohm's law since it does not involve the electrical field $\mathcal{E}_x$. In this case, the true electromotive force is given by the gradient of the electrochemical potential as carriers experience both diffusion and effects of the electric field.

The simple derivation of Eq.~(\ref{eq:derivdiffTE2}) has been allowed by the use as a reference of a fixed energy level, arbitrarily chosen but independent of the position along the material, rather than the bottom of the conduction band. This approach demonstrates that a particular knowledge of both $\nabla_x n$ and $\mathcal{E}_x$ is not mandatory to determine the Seebeck coefficient of a non-degenerate semiconductor. Indeed, in this case, these two contributions to the electrochemical gradient seem to always compensate each other in such a way that the resulting electromotive power is independent of specific assumptions, for example a constraint on the carrier concentration. Equation~(\ref{eq:derivdiffTE2}) is thus valid for a wide range of temperatures: It is correct for the extrinsic regime, i.e., when the carrier concentration is fixed by the concentration of impurities, but it also remains valid in the freeze-out regime and in the intrinsic regime where additional carriers are thermally generated. However, in this latter regime, the electron hole contribution to thermoelectric power should also be considered as these minority carriers may no longer be negligible. One may also refer to Ref.~\cite{Rose1957} in which the authors derive the Seebeck coefficient focusing only on potentials and the electric field rather than using a statistical approach.

\subsection{Link with non-equilibrium thermodynamics}

While Eq.~(\ref{eq:TEmicro}) is widely used in solid-state physics, its formulation is slightly different in non-equilibrium thermodynamics as general forces are traditionally computed from the gradients $ \frac{1}{T}\nabla_x \widetilde{\mu}$ and $\nabla_x \left(1/T\right)$, instead of $\nabla_x \widetilde{\mu}$ and $\nabla_x T$ \cite{Callen1985,Goupil2011}. So, one should then rewrite Eq.~(\ref{eq:TEmicro}) to get:

\begin{equation}\label{eq:thermodynamics}
J_x = \frac{\sigma T}{e} \cdot \frac{1}{T}\nabla_x \widetilde{\mu} + \sigma \alpha  T^2 \cdot \nabla_x \left(\frac{1}{T}\right).
\end{equation}

\noindent With this form, it is possible to identify each term with the canonical expression \cite{Callen1985},

\begin{equation}\label{eq:Callen}
- J_N = \frac{J_x}{e} = L_{11} \cdot \frac{1}{T}\nabla_x \widetilde{\mu} + L_{12} \cdot \nabla_x \left(\frac{1}{T}\right),
\end{equation}

\noindent to recover the expressions of the kinetic coefficients in the thermoelectric case, i.e., $L_{11} = \sigma T/e^2$ and $L_{12} = \sigma \alpha  T^2 / e$.

\subsection{On the so-called effective Seebeck coefficient}
Let us now turn to the previous analysis of the thermoelectric power in non-degenerate semiconductor. In ref. \cite{Mahan2000}, Mahan introduces an effective Seebeck coefficient $\overline{S}$, distinct from the genuine thermoelectric power eq. (\ref{def1}) obtained from measurements. In a subsequent article with Cai \cite{Cai2006}, this effective coefficient is presented as the ratio between the electric field and the temperature gradient. These two different Seebeck coefficients are related through the following relation~\cite{Cai2006}: 

\begin{equation}\label{relationSeebecks}
\alpha = \overline{S} + \frac{1}{e}\left(\frac{\partial\mu}{\partial T}\right)_n ,
\end{equation}

\noindent A comparison of eq. (\ref{eq:Stratton}) with the equation~(20) of ref. \cite{Mahan2000} leads to identify the effective Seebeck coefficient to the contribution of the diffusivity gradient, i.e., $\alpha_D$. The second term of the right hand side of eq. (\ref{relationSeebecks}) should then be associated with the assumption of constant diffusivity, i.e., to eq. (\ref{alphaSC}). This latter term is identical to the so-called \emph{Kelvin formula of the thermopower} \cite{Peterson2010}. As discussed by Shastry \cite{Shastry2013}, this contribution ``\emph{captures the many body density of states enhancements, while missing velocity and relaxation contributions}''. It thus justifies the introduction of the coefficient $\overline{S}$ to take into account dynamical effects. We believe however that this coefficient should not be presented as \emph{effective} since it does not reflect the appearance of the electromotive force due to the temperature gradient. It represents only one of the possible contributions to this electromotive force. From a practical viewpoint, it has recently been demonstrated that the contribution to the thermopower from the diffusivity gradient might be significant \cite{Sun2015}.

\subsection{On the contact potentials}
Finally, we want to point out the inappropriate use of the contact potentials in the derivation of the thermoelectric power sometimes found in the literature. For example, in Ref.~\cite{Ioffe1960}, Ioffe obtains Eq.~(\ref{eq:derivdiffTE2}) splitting the Seebeck coefficient into three separate terms, one being $\alpha_D$ while the two others, $\alpha_n$ and $\alpha_\varphi$, are associated respectively to concentration gradient and to the ``temperature dependence of the contact potential''. However, as demonstrated later by Chambers \cite{Chambers1977}, contact potentials are irrelevant to thermoelectric effects. This latter term is indeed introduced only to compensate the erroneous expression of $\alpha_n$ stemming from the confusion between $\varphi$ and $\widetilde{\mu}/e$. From an experimental viewpoint, contact potentials are irrelevant since they cannot be probed by a voltmeter: As depicted in Fig.~\ref{fig:figure3}, these energy discontinuities concern only the bottom of the conduction band $E_C$ (or identically the vacuum level just outside the material $\epsilon_S$) but not the Fermi level $E_F$.

\section{Conclusion}
In this article, we have discussed the definition of the thermoelectric power with a special emphasis on its relationship to the electrochemical potential. A proper consideration of all potentials inside the material has led to demonstrate that the phenomenological equation for the electrical current involving thermoelectric coefficients may be derived directly from the drift-diffusion equation. We also shed light on the physical interpretation of the effective Seebeck coefficient defined by Mahan, showing that it is actually related to the gradient of diffusivity along the system.

\end{document}